# Some lattice models with hyperbolic chaotic attractors

## S. P. Kuznetsov[1,2]

[1] *Udmurt State University, Izhevsk, Russia*
[2] *Kotelnikov's Institute of Radio-Engineering and Electronics of RAS, Saratov, Russia*
spkuz@yandex.ru

Examples of one-dimensional lattice systems are considered, in which patterns of different spatial scales arise alternately, so that the spatial phase over a full cycle undergo transformation according to expanding circle map that implies occurrence of Smale – Williams attractors in the multidimensional state space. These models can serve as a basis for design electronic generators of robust chaos within a paradigm of coupled cellular networks. One of the examples is a mechanical pendulum system interesting and demonstrative for research and educational experimental studies.

## 1. Introduction

The uniformly hyperbolic attractors were introduced in mathematical theory of dynamical systems due to Smale, Anosov, Sinai, and other researchers in the 1960s – 1970s [1]. Hyperbolic attractors are characterized by roughness or structural stability. In the context of physical or technical objects it implies insensitivity of the dynamical behavior to small variations in parameters, manufacturing imperfections, interferences, etc. that may be significant for possible applications [2]. It turned out, however, that hyperbolic chaos is not widespread in real-world systems, and its implementation requires special efforts.

Smale – Williams solenoid is the simplest representative of the hyperbolic chaotic attractors. Imaging an abstract discrete-time dynamical system assume that the evolution in one step transforms certain torus region in such way that it experiences longitudinal stretching and tranversal compression, folds into a loop with the number of turns $M \geq 2$, and is placed inside the original torus. Under multiple repetitions, the number of turns tends to infinity, and the resulting object will have transversal Cantor structure. The essential point is that the angular coordinate evolves according to expanding circle map. For individual orbits on the attractor the dynamics are chaotic. For clarity, the above description appeals to the three-dimensional state space, but such attractors can occur in spaces of higher dimension too.

Physical examples of systems with attractors of Smale – Williams type can be constructed using oscillators residing in states of excitation and inhibition alternately, while the angular variable has a sense of the oscillator phase [3]. Another approach is based on treatment of patterns arising in an active medium, say, that for Turing structures or standing waves, and the angular variable is a spatial phase [4, 5, 6]. A disadvantage of the first approach is that it requires, as a rule, a use of rather complex external driving for parameter modulation, combining low-frequency and high-frequency components. Within the second approach, instead of high-frequency modulation a spatial non-homogeneity is introduced that is effortless. A disadvantage is a need of exploiting systems of infinite dimension of the state space, which complicates mathematical description and practical implementations.

An interesting seems application of the second approach for organizing hyperbolic chaos in finite-dimensional systems, namely, in lattices of cells, whose dynamics are governed by ordinary differential equations. In this concern, it is worth mentioning a paradigm of cellular neural networks (CNN) on a base of electronic components designed as arrays of cells arranged in space [7]. These systems were suggested, particularly, for parallel data processing, as an alternative to traditional computational approaches. As one of the directions, application of CNN was considered for analog modeling of complex space-time dynamics including Turing structures, spiral patterns, turbulence.

This article discusses three models of one-dimensional cell arrays, which can inspire design of CNN generating rough chaos.

## 2. Nonautonomous lattice system generating Turing patterns

The simplest non-autonomous lattice system can be obtained by spatial discretization of the model based on the Swift – Hohenberg equation [4], where alternate excitation of long-wave and short-wave Turing patterns is provided. Replacing the spatial differentiation operator in equation of Ref. [4] with a difference operator, we obtain

$$\dot{u}_j + 2\kappa^2(1+2\kappa^2)(u_{j-1} - 2u_j + u_{j+1}) + 4\kappa^4(u_{j-2} - 2u_j + u_{j+2}) = (A - 1 + \varepsilon\delta_j)u_j - u_j^3, \quad (2.1)$$

where $u_j$ is dynamical variable related to the $j$-th spatial cell, $\kappa$, $A$, $\varepsilon$ are parameters. The quantities $\delta_j$ define the spatial non-homogeneity, the role of which will be explained below. Note that in the lattice a connection is involved not only between the nearest neighbors but also with the neighbors through one, with a certain ratio between the coupling coefficients.

Let the system be a ring chain with the number of cells $2N$. If the parameter $\kappa$ is constant and $\varepsilon=0$, then in the linear approximation the increment of the pattern with the wavenumber $k$ is $\lambda = A - (1 - 4\kappa^2 \sin^2 \pi k/2N)^2$, and it is maximal at $\sin(\pi k/2N) = 1/2\kappa$.

In the ring with periodicity condition $u_{j+2N} = u_j$, the wave number $k$ should be integer. For patterns of wave numbers $k=1$ and $k=3$ the maximal increment is achieved, respectively, at $\kappa_1 = \frac{1}{2}\sin\frac{\pi}{N}$ and $\kappa_3 = \frac{1}{2}\sin\frac{3\pi}{N}$. We will assume that in equation (2.1) the parameter is modulated in time with period $T$, in such way that the alternate excitation of these patterns is provided: $\kappa(t) = \kappa(t+nT) = \begin{cases} \kappa_1, & 0 \leq t < T/2, \\ \kappa_3, & T/2 \leq t < T. \end{cases}$

At the stage $\kappa = \kappa_1$, a pattern is formed with the wave number $k=1$ and some spatial phase $\varphi$, i.e. $u \sim U_1 \cos(\pi j/N + \varphi) + \widetilde{U}_3 \cos(3\pi j/N + 3\varphi)$, where $U_1 \sim \sqrt{A}$. The third harmonic arise due to the presence of a cubic term and has a small amplitude $\widetilde{U}_3 \ll U_1$. After switching to $\kappa = \kappa_3$, the long-wave component $k=1$ decays, but the system becomes unstable with respect to the excitation with $k=3$. Initial stimulation of the short-wave pattern is provided by $\widetilde{U}_3$, so that it accepts a spatial phase of $3\varphi$. At the end of the stage the pattern $u \sim U_3 \cos(3\pi j/N + 3\varphi)$ takes place, with $U_3 \sim \sqrt{A}$. After the next switching, the third harmonic decays, but instability for the first harmonic gives rise to growth of its amplitude. The "germ" is the component with $k=1$ provided by combination of the damping short-wave mode and the spatial distribution $\varepsilon\delta_j$. If the fourth harmonic dominates in $\delta_j$ then, due to the term $\cos(4\pi j/N)\cos(3\pi j/N + 3\varphi) = \frac{1}{2}\cos(\pi j/N - 3\varphi) + ...$, the long-wave pattern arises, this time with the phase of $-3\varphi$. Thus, at each full modulation period, the spatial phase is transformed in accordance with a three-fold expanding circle map.

Taking into account that spatial distributions of $u_j$ are determined by odd harmonics, we can consider a lattice of twice less number of cells $N$. To do so we simply replace the boundary conditions of periodicity by the condition of sign change: $u_{-1} = -u_{N-1}$, $u_N = -u_0$. One can use a stroboscopic description of the dynamics on a period of the modulation by the Poincaré map $\mathbf{X}_n = \mathbf{F}(\mathbf{X}_{n-1})$, where the state vector is $\mathbf{X}_n = (u_0, u_1, ..., u_{N-1})_{t=nT}$.

Figure 1*a* illustrates space-time dynamics in the lattice system with $N=12$. Distributions of the values $u_j$ are shown at the instants of the parameter switches depending on the spatial index. Although $u_j$ relate to discrete nodes, the points are linked in the plot to visualize the patterns clearly. As seen, the waveforms at each new stage of activity jump chaotically over the lattice length. It can be verified that the chaotic displacement of the patterns at successive stages corresponds to transformation of the spatial phases according to the triple expanding circle map. During numerical integration of the equations, at the end of each modulation period a spatial phase of the pattern is evaluated as $\varphi_n = \arg(u_0 + iu_{N/2})$, and the data are plotted in coordinates $(\varphi_{n-1}, \varphi_n)$, see panel *b*. Observe that one passage of the full interval for the pre-image corresponds to a triple passage of the image in the opposite direction. Compression of the phase volume in other directions in the state space ensures existence of the Smale – Williams type attractor. Panel *c* shows a set of points $(u_0, u_{N/2})$ obtained by successive iterations of the stroboscopic Poincaré map. An enlarged fragment of the plot illustrates the transversal Cantor structure

inherent to the solenoid.

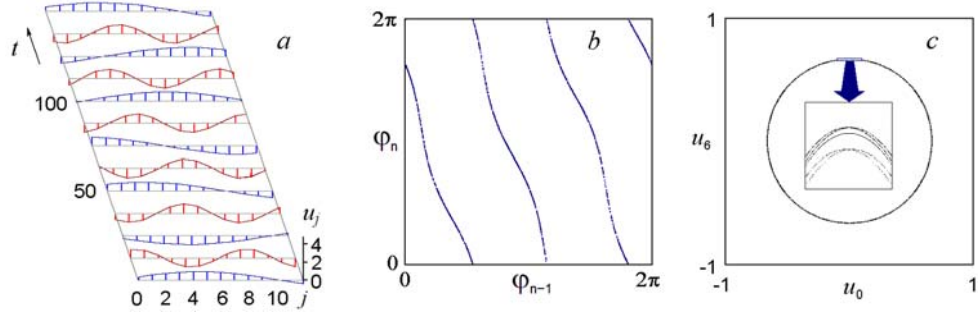

Fig.1. (*a*) Evolution of patterns in the ring system (1) with the number of cells *N*=12 and boundary condition of the sign change at the ends. The configurations relating to the parameter switching instants are shown. (*b*) Diagram of the spatial phase transformation on each one parameter modulation period. (*c*) Portrait of the attractor of the Poincaré map in projection on the plane. Parameters are: *A*=0.4, *T*=25, and ε=0.03 with the spatial non-homogeneity imposed with the dominated fourth harmonic: $\delta_j = \{0,1,1,0,-1,-1,0,1,1,0,-1,-1\}$.

Total number of Lyapunov exponents for the system is *N*=12, however, to judge on the nature of the attractor and its fractal properties it is sufficient to evaluate only few larger exponents. Numerically, the first three exponents at the chosen parameters are $\Lambda_1 = 0.993$, $\Lambda_2 = -5.085$, $\Lambda_3 = -34.19$. The positive one is responsible for the instability of trajectories on the attractor and for chaotic nature of the dynamics. It is close to the value $\ln 3 \approx 1.0986$ associated with the expanding map for the angular variable. The remaining negative exponents are responsible for the transverse compression and formation of the Cantor structure of the solenoid. Using the Kaplan – Yorke formula, one can estimate the fractal dimension of the solenoid as $D_{KY} = 1 + \Lambda_1 / |\Lambda_2| \approx 1.20$.

## 3. Pendulum ring chain with vibrating suspension

A pendulum, the suspension point of which performs a periodic oscillatory motion in the vertical direction, is an interesting example of a mechanical system, where, depending on parameters, many different modes can be observed [8]. The phenomenology of dynamics becomes even richer if we turn to systems based on chains of coupled pendulums and to continuous limit that is a medium described by the sine-Gordon equation [9].

Here we will consider a set of *N* pendulums suspended on a ring hoop, which is forced to perform a definite oscillatory motion in the vertical direction. Each pendulum is connected with the nearest neighbors by spiral springs, so that the moment of the interaction force is proportional to a relative deflection angle of the neighboring pendulums. Avoiding unnecessary complication of the model, we assume that dissipation is due to the presence of friction force between the hoop and the pendulums of moment proportional to the instantaneous angular velocity. The equations in dimensionless form are

$$(1+\varepsilon\delta_j)[\ddot{\theta}_j + (1+a(t))\sin\theta_j] = -\gamma\dot{\theta}_j + D(\theta_{j-1} - 2\theta_j + \theta_{j+1}), \; j=0,1,...,N-1, \quad (3.1)$$

where $\theta_j$ is a deflection angle for the *j*-th pendulum, $\gamma$ is dissipation parameter, *D* is coupling coefficient of the neighboring pendulums, $\varepsilon\delta_j$ is relative deflection of a mass of the *j*-th pendulum from the mean value. The ring arrangement of the pendulums implies imposition of the boundary conditions of periodicity: $\theta_{j+N} = \theta_j$.

Let the vertical movement imparted to the hoop, to which the pendulums are suspended, follows a sinusoidal law with amplitude $A_2$ and frequency $\omega_2$ during $N_2$ oscillations, and then with amplitude $A_1$ and frequency $\omega_1$ during $N_1$ oscillations, after which the switches are repeated with period *T*:

$$a(t) = \begin{cases} A_2\omega_2^2 \sin\omega_2 t, \; 0 \le t < \tau, \\ A_1\omega_1^2 \sin\omega_1(t-\tau), \; \tau \le t < T, \end{cases} \quad \tau = 2\pi\frac{N_2}{\omega_2}, \; T = 2\pi\left(\frac{N_2}{\omega_2} + \frac{N_1}{\omega_1}\right). \quad (3.2)$$

Bearing in mind that the eigenfrequencies of linear modes of oscillations of the ring chain of pendulums without taking into account pumping, dissipation and mass variation, are given by $\Omega_s = \sqrt{4D\sin^2 \pi s N^{-1} + 1}$, we assign the pump frequencies $\omega_1$ and $\omega_2$ equal to twice the frequencies of the first and the third modes: $\omega_1=2\Omega_1$, $\omega_2=2\Omega_3$.

The dynamics may be treated in terms of stroboscopic Poincaré map $\mathbf{X}_n = \mathbf{F}(\mathbf{X}_{n-1})$ transforming 2*N*-dimensional state vector $\mathbf{X}_n = (\theta_0, \theta_1,...,\theta_{N-1}, \dot{\theta}_0, \dot{\theta}_1,...,\dot{\theta}_{N-1})_{t=nT}$ for one period of the pump modulation.

The mechanism leading to emergence of the hyperbolic chaos is similar to that for the model of parametric excitation of oscillations of a nonlinear string of Ref. [5]. During pumping at the frequency $\omega_1$ a standing wave is excited, for which we can write, roughly, $\theta_j \sim \sin(2\pi j/N + \varphi)$, where the phase constant $\varphi$ depends on initial conditions. The amplitude stabilizes at some level due to nonlinearity of the pendulums. Also, because of the nonlinearity, there will be a component of the third spatial harmonic, having a spatial phase of 3$\varphi$. At the next stage, the pumping at $\omega_1$ stops, and the oscillations of the first mode are damped. Now, however, pumping at frequency $\omega_2$ is switched on, which leads to parametric instability for standing wave of the third mode. This wave is formed from the initial perturbation given by the third spatial harmonic of the wave produced at the previous stage, so it will have the spatial phase shift of 3$\varphi$. Further, the stage of pumping at $\omega_1$ comes again. The seed for the parametric oscillations growth is provided by a combination of the disturbance $\theta_j \sim \sin(6\pi j/N + 3\theta)$ left from the previous stage and the spatially non-uniform mass distribution given by $\delta_j$. If it contains the dominating second harmonic $\delta_j \sim \sin(4\pi j/N)$, the combination contributing to the first mode has a spatial phase $\varphi_{new} = 3\varphi + \text{const}$, as $\sin(4\pi j/N)\sin(6\pi j/N + 3\theta) = \frac{1}{2}\cos(2\pi j/N + 3\theta) + ...$, and the parametrically excited standing wave will inherit the same phase. Thus, at each new modulation period, a three-fold expanding circle map for the phase takes place. Result of multiple repetition of the transformation will be formation of the Smale – Williams solenoid in the state space of the map $\mathbf{X}_n = \mathbf{F}(\mathbf{X}_{n-1})$.

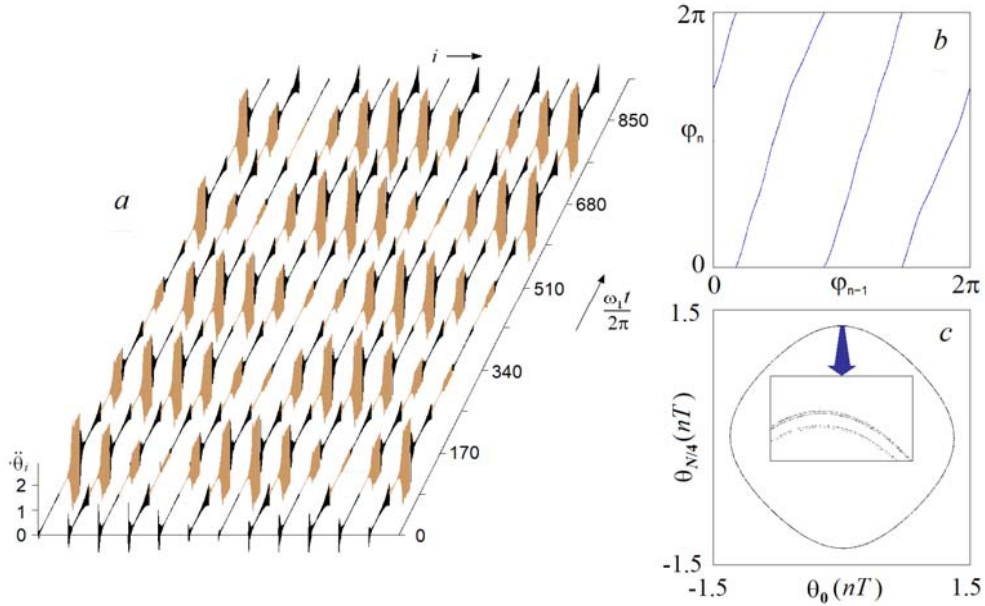

Fig.2. (*a*) Diagrams of dependences of the angular accelerations of pendulums on time in sustained chaotic mode basing on numerical solution of the differential equations, where the stages of slow and fast oscillations of the suspension are marked with brown and black. (*b*) Plot of the spatial phase transformation on each one modulation period of pumping. (*c*) Portrait of the attractor of the Poincaré map in projection on the plane. The number of chain elements is $N=12$, parameters are $D=1.19$ and $\gamma=0.12$, $A_1\omega_1^2 = A_2\omega_2^2 = 0.6$. The variation of the masses is characterized by $\varepsilon=0.01$ and a set $\delta_j=\{0, 1, 1, 0, -1, -1, 0, 1, 1, 0, -1, -1\}$. Switches of the pump frequencies between $\omega_1=2.297$ and $\omega_2=3.677$ take place after each $N_1=85$ and $N_2=136$ oscillations of the suspension.

Figure 2*a* shows dependences of the angular accelerations of the pendulums on time according to data of numerical integration of (3.1) for sustained chaotic motion. Although pendulum oscillations themselves are indistinguishable on the scale of the graph, it shows clearly how the amplitudes vary in time in the system functioning in accordance with the above described mechanism. Panel *b* illustrates transformation of spatial phase of the standing waves during each modulation period of the pump. The phases $\varphi_n = \arg[\theta_0(nT) + i\theta_{N/4}(nT)]$ are calculated at the moments of switching the pump frequency from $\omega_1$ to $\omega_2$, when the first spatial mode is dominating. This diagram is the main evidence that the Smale – Williams type attractor indeed takes place since it demonstrates the required topological property. In the system under consideration, the solenoid is an object in the state space of the Poincaré map of dimension 2*N*=24. Panel *c* shows this attractor in two-dimensional projection. The enlarged fragment in the diagram visualizes the transverse structure of fibers of the solenoid.

A total number of Lyapunov exponents of the attractor of the Poincaré map is 2*N*=24. The first three exponents are $\Lambda_1 = 1.0913, \Lambda_2 = -2.484, \Lambda_3 = -14.70$. A positive exponent responsible for chaotic nature of the dynamics is close to the value $\ln 3 \approx 1.0986$ associated with the threefold expanding map. The remaining exponents are negative being responsible for the transversal compression and formation of the Cantor structure of the solenoid of fractal dimension estimated from the Kaplan – Yorke formula as $D_{KY} \approx 1.44$.

## 4. Autonomous lattice system with hyperbolic chaos

In this Section an autonomous lattice system is examened inspired by a distributed system with Smale – Williams attractor proposed in Ref. [6]. We change the spatial differentiation by finite-difference operators and add some modifications to adopt the model to simpler implementation. Consider a ring chain of 2*N* cells governed by equations

$$\dot{u}_j = D_0(u_{j-1} - 2u_j + u_{j-1}) + u_j^3 - u_j v_j^2 - \alpha u_{j+N} + \varepsilon \delta_j v_j, \quad \dot{v}_j = (-\gamma + u_j^2) v_j + \mu u_j^2. \quad (4.1)$$

Here $u_j, v_j$ are the dynamic variables of the cells numbered by $j = 0,1,...2N-1$, and μ, ε, γ are parameters. In addition to coupling of neighboring cells characterized by $D_0$, each cell is linked with the opposite element of the ring that is characterized by coefficient α. By a set of values $\delta_n$, a weak spatial inhomogeneity is introduced.

If ε=0, then, near the equilibrium state the substitution $u_j \sim \exp(\lambda t - i\pi k j/N)$ leads to expression for the increments of modes of wave numbers *k*, which, in accordance with the condition of periodicity, should be integers: $\lambda(k) = -\alpha(-1)^k - 4D_0 \sin^2(\pi k/2N)$. Assuming $0 < \alpha < 4D_0 \sin^2(3\pi/2N)$, only one mode *k*=1 has a positive increment while the other modes are damped, including the homogeneous one with *k*=0.

When specifying the spatial non-homogeneity as the third spatial harmonic, i.e. $\delta_j \sim \cos 3\pi j N^{-1} + ...$, it is possible to ensure generation of the hyperbolic chaos.

Suppose the system initially is close to zero state and demonstrates growth in time of the spatial distribution $u_j$ with the wave number *k*=1. In general case, this is a superposition of the sine and cosine components with some coefficients that can be written as a single term with some spatial phase φ: $u_j \sim \cos(\pi j N^{-1} + \varphi)$. When the factor $(-\gamma + u_j^2)$ in the second equation becomes positive, the growth of the variable $v_j$ starts. Since at its initiation the process is stimulated by a quadratic term $u_j^2$, the spatial dependence of $v_j$ will be determined by the second harmonic: $v_j \sim \cos^2(\pi j N^{-1} + \varphi) = \tfrac{1}{2} + \tfrac{1}{2}\cos(2\pi j N^{-1} + 2\varphi)$. As the values of $v_j$ grow, the variables $u_j$ start to decrease rapidly at some time due to the inhibitory effect of the term $v_j^2$ in the first equation. When the values $u_j$ become small enough, the variables $v_j$ also experience damping with decrement determined by γ. As a new stage of increase of $u_j$ comes, it is stimulated by a term $\varepsilon \delta_j v_j$, namely, by its first harmonic: $\delta_j v_j \sim \cos(3\pi j/N)\cos(2\pi j/N + 2\varphi) = \tfrac{1}{2}\cos(\pi j/N - 2\varphi) + ...$ This ensures transfer of the double phase with the opposite sign to the first harmonic of $u_j$: $\varphi_{new} \approx -2\varphi$.

Similar dynamics can be provided in a lattice of twice less number of cells. Indeed, as the distributions of $u_j$ are determined by odd harmonics, and for $v_j$ by even ones, the solutions of (4.1) we consider must satisfy $u_{j+N} = -u_j$ and $v_{j+N} = v_j$. Therefore, instead of (4.1) we can write the equations for *N* cells

$$\dot{u}_j = D_0(u_{j-1} - 2u_j + u_{j+1}) + u_j^3 - u_j v_j^2 + \alpha u_j + \varepsilon \delta_j v_j, \quad \dot{v}_j = (-\gamma + u_j^2)v_j + \mu u_j^2, \quad (4.2)$$

where the boundary conditions of periodicity are replaced by conditions of sign change at the ends: $u_{-1} = -u_{N-1}, u_N = -u_0$. Here, unlike (4.1), the long-distance interactions are excluded; the respective term in the equation contains now the variable for the same cell with inverted sign.

The system (4.2) can also be described in terms of the Poincaré map, the dimension of which is $2N-1$. To do so, we need to introduce a cross-section in the state space of the system (4.2) by some hypersurface *S*, which is defined by some algebraic equation $f(u_0, v_0, ..., u_{N-1}, v_{N-1}) = 0$ and must cross the flow of phase trajectories. The Poincaré map expresses a vector of a point on the hypersurface through the vector of the previous its intersection by the trajectory: $\mathbf{X}_n = \mathbf{F}(\mathbf{X}_{n-1})$.

Figure 3 shows space-time diagrams illustrating dynamics of the ring system (4.1) with the number of cells 2*N*=12. The distributions of the values *u* and *v* are shown depending on the spatial index *j* plotted along the horizontal axis at time instants corresponding to the maximal values of the first mode amplitude. As seen from the figure, the waveforms at each new stage of activity jump chaotically on the chain length.

Similar dynamics occur for the same parameters in the system (4.2) with the number of cells *N*=6 with the boundary conditions of sign reversal at the ends, see Fig. 4. The chaotic displacement of the patterns at successive stages of activity corresponds to transformation of the spatial phases according to the double expanding circle map. In the process of numerical integration of the equations at a time of each *m*-th maximum of the first mode amplitude evaluated as $\sqrt{u_0^2 + u_{N/2}^2}$, the spatial phase is calculated as $\varphi_n = \arg(u_0 + i u_{N/2})$, and the data are plotted in coordinates $(\varphi_{n-1}, \varphi_n)$, see panel *b*. Although the shape of the branches is distorted in comparison with the ideal linear expanding circle map, this does not violate its affiliation to the same topological class. In presence of the phase volume compression in other directions in the state space it ensures occurrence of the Smale – Williams type attractor. Panel *c* shows set of points ($u_0$, $u_{N/2}$) corresponding to the moments of maximal amplitudes of the first mode. This is a portrait of the attractor in Poincaré section in a two-dimensional projection.

The total number of Lyapunov exponents for the system (4.2) is 2*N*=12. The computed first three exponents are $\lambda_1 = 0.0597$, $\lambda_2 = 0.0000$, $\lambda_3 = -0.2046$. As the average period of the Poincaré section passages is $T \approx 9.866$ according to the calculations, the largest Lyapunov exponent of the Poincaré map is $\Lambda_1 = \lambda_1 T \approx 0.589$ that roughly agrees with the value $\ln 2 \approx 0.693$. The second exponent is zero and refers to a perturbation along the reference trajectory in the autonomous system. The remaining exponents are negative. Kaplan – Yorke dimension for the Poincaré map attractor is $D_{KY} = 1 + \lambda_1 / |\lambda_3| \approx 1.29$ that reflects the fractal structure of the solenoid.

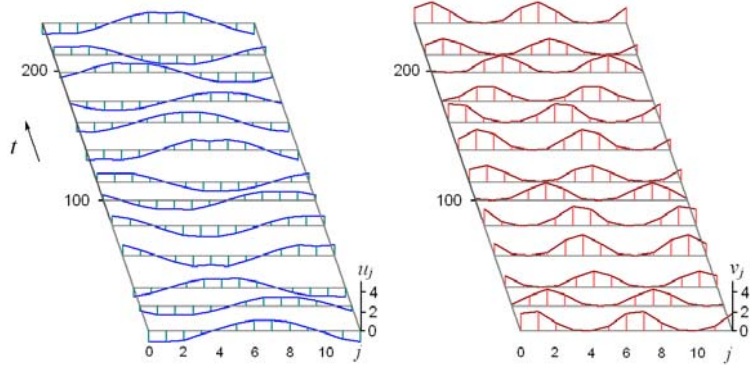

Fig.3. Evolution of patterns in the ring system (4.1) with the number of cells 2*N*=12 in the sustained chaotic regime, on the left for the variable *u*, and on the right for the variable *v*. The configurations relating to successive maximums of the first mode amplitudes are shown. Parameters are: $D_0$=8, α=2.2, γ=0.6, μ=0.4, ε=0.25, $\delta_j = \{1,-1,-1, 1, 1, -1,-1, 1, 1,-1,-1, 1\}$.

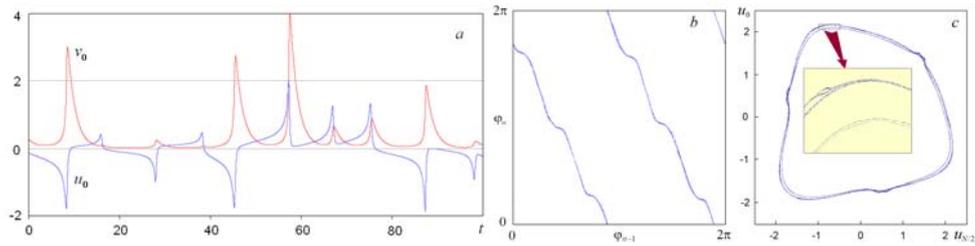

Fig.4. Waveforms in the sustained regime of chaotic self-oscillations for a cell *j*=0 in system (4.2) (*a*), diagram for the spatial phases (*b*) and attractor in the Poincaré section (*c*). Parameters are $D_0$=8, α=2.2, γ=0.6, μ=0.4, ε=0.25, $\delta_j = \{1,-1,-1, 1, 1, -1\}$.

## 5. Conclusion

We considered three lattice models in a form of one-dimensional arrays of coupled cells, which are able to generate hyperbolic chaos associated with attractors of Smale – Williams type. These examples show a possibility of implementing such attractors in finite-dimensional systems, where a spatial phase of forming and disappearing patterns undergoing expanding circle map on a characteristic time interval plays a role of angular variable on Smale – Williams solenoids. The models, as it may be expected, can inspire design of electronic generators of rough chaos, following the earlier suggested paradigm of coupled neural networks. One of the examples is a ring pendulum system, which seems interesting and demonstrative mechanical model for research and educational experimental studies illustrating the hyperbolic chaos.

## Funding

The work was partially supported by Russian Science Foundation grants no. 17-12-01008 (Section 2) and 15-12-20035 (Section 3).